# Simulation model of spacetime with the Minkowski metric

Vasyliy I. Gurianov, e-mail: vg2007sns@rambler.ru

**Abstract.** *In this paper, a simulation model of spacetime with the Minkowski metric is proposed. Object-oriented simulation (OOS) is used for simulation. Ontology of Special Relativity Theory (SRT) is proposed, as well as a mechanism for the synchronization of laboratory time and local time. This paper also presents the main experiments conducted to demonstrate relativistic effects. Differences between simulation and numeric modelling are discussed.*

**Keywords:** simulation, special relativity, discrete spacetime, energy-momentum, UML2 SP

## 1. Introduction

The use of simulation in such innovative fields such as nanotechnology is of great practical and scientific interest. However, today, as a rule, numerical modelling is used. This is due to the problem of the simulation of relativistic and quantum effects.

In this paper, we propose a simulation model of spacetime with the Minkowski metric. We use object-oriented simulation (OOS), but the model is also suitable for agent-based simulation (ABS).

The idea that spacetime is a computational process was first put forward by Konrad Zuse in the book [1] in 1967.

Stephen Wolfram [2, 3] suggested that fundamental physical laws could be derived from cellular automata, and, in particular, the properties of spacetime. In his opinion, space is a network, i.e. graph. The well-known properties of space appear in the same way as statistical physics transforms into thermodynamics. That is, this is a result of the appearance of a large number of nodes and links. Time appears as a result of a change in the state of state machines, but not as a spacetime network.

Gerard't Hooft is of a similar opinion. His point of view is outlined in the book [4]. Superdeterminism allows us to consider nature as the result of the operation of cellular automatons. Quantum effects arise from the use of templates in macroscopic measurements. Gerard't Hooft does not directly consider spacetime, but he demonstrates how one can arrive at classical physics.

Our approach is different in that we consider spacetime as an object of simulation. The simulation model is described in language UML2 SP. We also use a spacetime discrete model, but it is the result of describing the model as a semantic net. We do not assert that spacetime is discrete.

This paper is organised as follows: Section 2 contains a brief description of the UML2 SP language. Section 3 formulates the model of spacetime. The main experiments on the simulation of relativistic effects are presented in Section 4. Finally, Section 5 presents the paper's conclusions.

## 2. About UML2 SP

The base conception of UML2 SP [5] is a communicative paradigm: any object of reality can be described as a system of elementary communicative acts.

In the UML2 SP, UML-diagrams have a double semantic. In the computational semantic, the class diagram is a model of the program. In the problem domain semantic, the class diagram is ontology. Class is considered as a frame as per Marvin Minsky. The name of the class is the name of the frame. The tagged value 'Concept' defines a notion from the problem domain. Attributes of the class are slots of the frame. Operations are procedures of the frame. Attributes and operations also define concepts.

The base method of UML2 SP is the decomposition principle, as in the IDEF0 methodology. That is, UML2 SP is an object-oriented variant IDEF0. The syntax of UML2 SP consists of fifteen stereotypes for the execution of decomposition.

The basic idea of the usage of UML2 SP is description of the object of study as a semantic net. The semantic net is the non-numeric model. Thus, UML2 SP models may be used for modelling objects with absent mathematical models.

## 3. Model of spacetime

### 3.1. Ontology of Special Relativity Theory

The model described in UML2 SP is an ontology.

The ontology of the Special Relativity Theory (SRT) is depicted in Fig.1. We use the C++ syntactic.

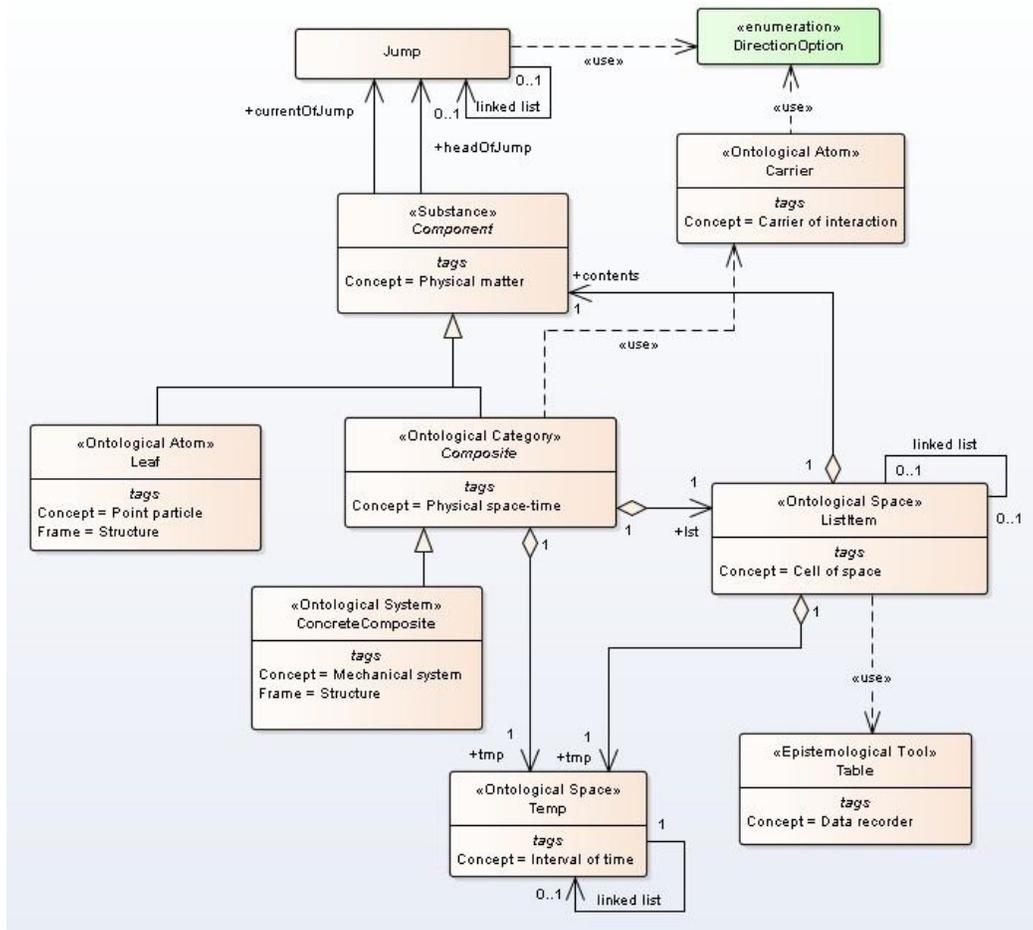

Fig. 1: Class diagram for the simulator of Special Relativity.

The 'Composite' frame defines the concept of 'Physical spacetime'. The 'tmp' attribute defines laboratory time. It is a linked list from the 'Temp' nodes. The 'lst' attribute defines one-dimension physical space. It is a linked list from the 'ListItem' nodes. The 'Leaf' frame defines the 'Point particle' concept. A particle may move in the space. The particle moves from one cell to another cell in 'lst'. The simulation is run by the 'Run()' operation of the 'ConcreteComposite' class.

Concepts in relativistic mechanics are similar to the concepts of classical mechanics, but there are differences. This is a synchronization rule of time of a mechanical system and local time of physical space cells.

### 3.2. Synchronization

The main concept in the STR-model is synchronization. We introduce two notions of time. Laboratory time is the time of the rest frame (attribute 'tmp' of the 'Composite' class, see Fig.1). Local time is time in each cell of physical space (attribute 'tmp' of the class 'ListItem', see Fig. 1). We must synchronize local time in each cell with laboratory time.

We define the 'Synchronization' concept through two notions: a synchronization mechanism and a synchronization rule.

The synchronization mechanism is depicted in Fig. 2.

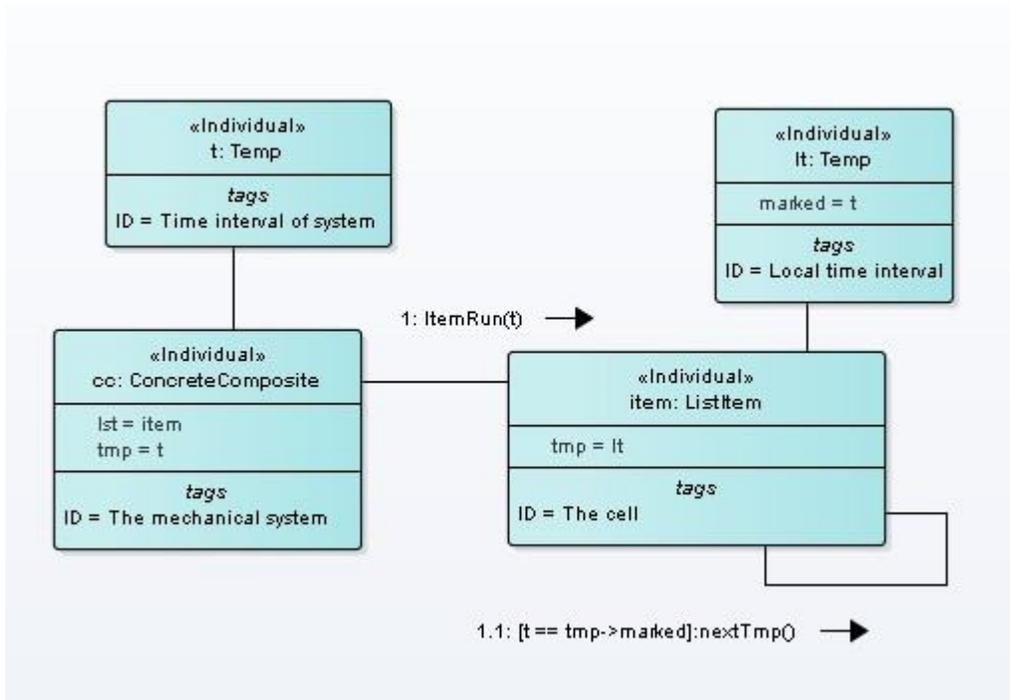

Fig. 2: Collaboration diagram for the synchronization mechanism.

Laboratory time is a 'tmp' linked list. The progress of the time is a node shift of the linked list. It is a main loop in the 'Run()' operation. The 'cc' object sends the message 'itemRun' to all 'item' objects. The message argues 't'. It is the current node of the 'tmp' linked list. The local time shift if 't' equals the 'marked' pointer.

We introduce also two new notions. If node t has t.lb = true, then it is called a 'bearing' time moment. All nodes of the 'tmp' linked list in 'ListItem' are 'bearing' moments. Nodes of the 'tmp' linked list in 'Composite' may be 'bearing' or 'ordinary' moments. Let $T_w$ be the number of ticks of model time. The numbers of nodes between 'bearing' moments is called the resolution of the tick of time ($\tau_R$).

The synchronization rule defines an assignment rule of a pointer for the 'marked' attribute in each cell of space. The synchronization rule is the following. We introduce the attributes 't' (type int) for Temp and 'x' (type int) for 'ListItem' for the measurement of time and coordinates. Let $\tau$ be a value of the variable 't', $\rho$ be value of the variable 'x', and $\sigma = \tau_R T_w$; then the synchronisation rule has the form of the invariant interval $\tau^2 = \sigma^2 + \rho^2(v_t/v_l)^2$, where $v_t$, $v_l$ are the coefficients of the conversion. We search for a node with number $\tau$ in the 'tmp' list. The node found is assigned the attribute 'marked'. This may be implemented with different algorithms. The natural algorithm for generating the temporal network is a separate problem.

Thus, 'Temp' and 'ListItem' instances form a network. The network is a simulation of spacetime with the Minkowski metric.

If time has shifted in a cell, then some activity in the cell is possibly.

### 3.3. Mechanical motion and interaction

In this section, we will define kinematic and dynamic concepts: the concept of 'mechanical motion' and 'interaction' concept.

We introduce two statements.

- A particle can move or interact.
- Interaction is possible if motion is completed.

The attributes 'headOfJump' and 'currentOfJump' define the 'Momentum' concept (for a single mass). The semantic net for the concept is depicted in Fig.3.

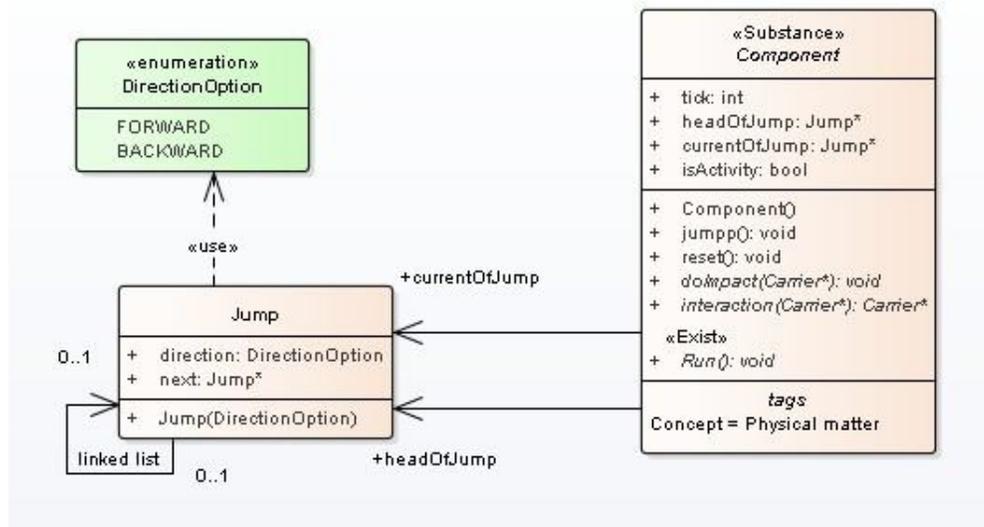

Fig. 3: Semantic net for the 'Momentum' concept.

The 'Mechanical motion' concept is defined through the mechanism of motion. If a particle in the cell and 'currentOfJump' is not NULL, then 'currentOfJump' has shift (message Run), and the particle moves to a neighbouring cell. If 'currentOfJump' is NULL, then the motion is completed. If the 'tmp' node is 'bearing' then the mechanical system sends the message 'reset' to the particle and the particle able to move again (currentOfJump is assigned the value headOfJump).

By $c$ denote light speed. Let $\tau$ be time, $\rho$ be distance, and $\mu$ be mass in natural units. Then, time $t$, distance $d$, and mass $m$ in standard units are calculated as

$$t = \tau/(cv_t),\ dim\ cv_t = L^{-1},$$
$$d = \rho/v_l,\ dim\ v_l = L^{-1},$$
$$m = \mu/v_m,\ dim\ v_m = M^{-1},$$

where $v_t$, $v_l$, $v_m$ are the coefficients of the conversion of time, distance, and mass.

Relative velocity is:

$$\beta = v/c = \frac{\rho}{\tau},$$

where $\rho$ - travel distance, $\tau = T_w \tau_R$, $T_w$ – number of bearing nodes.

The 'Interaction' concept is defined as 'doImpact' process. The operation has an argument of 'Carrier' type. The operation increases or reduces the 'headOfJump' linked list.

Let $\iota_i$ be the number of interaction acts in tick $i$ (interaction intensity). Force $f_i$ is

$$f_i = \frac{1}{\upsilon_m} \frac{\upsilon_t}{\tau_R^2} \iota_i.$$

The output of the formula may be found in documentation [6].

Let $j$ be the length of the 'headOfJump' list, $j_c$ is segment length of the 'currentOfJump'. The particle may not have a velocity of more than light speed as the $j_c$ is not more than $\tau_R$ even if $j > \tau_R$. The particle receives the message 'reset' if 'tmp' is bearing node. If $j = \tau_R$, then particle time stops and interaction is impossible. If $j = 0$, then time and interaction are as in classical mechanics. If $0 < j < \tau_R$ then particle time is slows down and intensity of interaction falls.

## 4. Results and discussion

### 4.1. Time dilation
Let us consider a motion with velocity of $\beta = 0.5$. In the table, the result of the simulation is depicted (units are metres and metres of light time).

Table 1: Trajectory of particle and time of particle

| Tw | x | t | ta | err% | tp |
|---|---|---|---|---|---|
| 0 | 0.0 | 0.0 | 0.0 | 0.0 | 0.0 |
| 1 | 0.5 | 1.2 | 1.12 | 7.33 | 1.0 |
| 2 | 1.0 | 2.3 | 2.24 | 2.86 | 2.0 |
| 3 | 1.5 | 3.4 | 3.35 | 1.37 | 3.0 |
| 4 | 2.0 | 4.5 | 4.47 | 0.62 | 4.0 |
| 5 | 2.5 | 5.6 | 5.59 | 0.18 | 5.0 |
| 6 | 3.0 | 6.8 | 6.71 | 1.37 | 6.0 |
| 7 | 3.5 | 7.9 | 7.83 | 0.94 | 7.0 |

Column Tw is the number of tick of model time. Column x – is the coordinate of the particle in moment Tw. Column tp is the time of the particle. We can observe dilation of time. In the particle, tp units of time elapse, but, in the laboratory frame of reference, t units of time are registered. The world line of the particle is depicted in Fig.4.

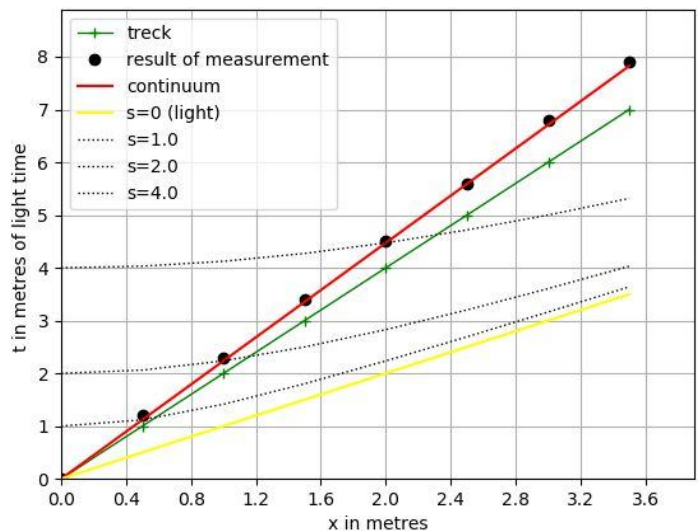

Fig. 4: Minkowski spacetime diagram for β = 0.5

## 4.2. Velocity, momentum, and energy

Let us consider the following characteristic experiment. Let a constant electric field act on a charged particle

$$\frac{dp}{dt} = qE_f.$$

We introduce new variables: $p \to pm_0c$, $E \to Em_0c^2$, where $c$ is the light speed, $p$ is the momentum, $E$ is the energy, and $m_0$ is the rest mass. We put $qE_f/(m_0c) = 0.1$ (one-act interaction), the rest mass of the particle is $m_0 = 1$ ('Skip' list is NULL), the particle initially rests at the origin of the laboratory reference frame. The results of the experiment are presented in Table 2.

Table 2: Velocity and energy of the particle as a function of momentum

| Tw | p | v | va | v,err% | E | Ea | E,err% |
|---|---|---|---|---|---|---|---|
| 0 | 0.0 | 0.0 | 0.0 | 0.0 | 1.0 | 1.0 | 0.0 |
| 1 | 0.11 | 0.09 | 0.11 | 16.86 | 1.01 | 1.01 | 0.39 |
| 2 | 0.21 | 0.2 | 0.21 | 2.68 | 1.03 | 1.02 | 0.8 |
| 3 | 0.31 | 0.3 | 0.3 | 1.32 | 1.06 | 1.05 | 1.25 |
| 4 | 0.42 | 0.36 | 0.39 | 6.09 | 1.1 | 1.08 | 1.42 |
| 5 | 0.53 | 0.45 | 0.47 | 2.94 | 1.15 | 1.13 | 1.61 |
| 6 | 0.64 | 0.55 | 0.54 | 1.19 | 1.21 | 1.19 | 1.91 |
| 7 | 0.76 | 0.58 | 0.61 | 3.59 | 1.28 | 1.26 | 1.91 |
| 8 | 0.88 | 0.67 | 0.66 | 0.91 | 1.36 | 1.33 | 2.1 |

The following notation is introduced in this table: Tw is the system time step number, p is the measured momentum, v is the measured velocity, va is the exact value of the velocity, v, err% is the relative error of the speed measurement in %, E is the measured energy, Ea is the exact value of energy, E, err% - is the relative error of the energy measurement in %.

Motion start has delay $\tau_d = \tau_R\,(\mu/\iota)$. It is caused by the appearance of energy of rest.

The plot of the dependence of velocity on momentum is presented in Fig. 5.

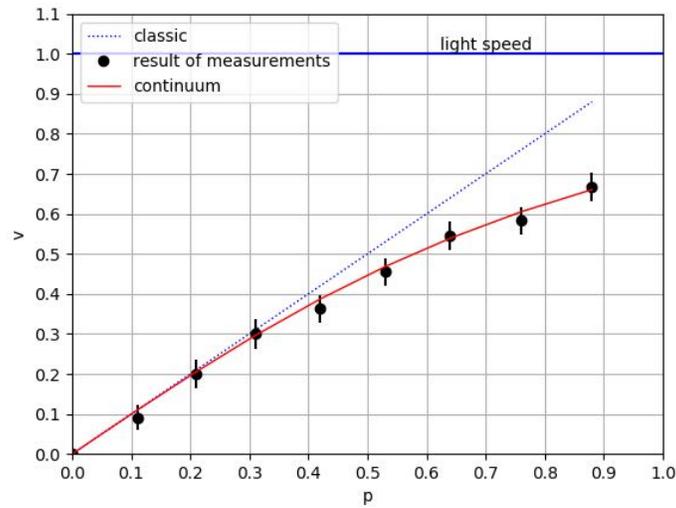

Fig. 5: Velocity as a function of momentum

These and other experiments may be found in our repository [6]. The repository contents code of the software (Python) and documentation.

### 4.3. Simulation and numerical modelling

We will explain our point of view on simulation and numerical modelling. Simulation models are used in science on occasion; for example, cellular automata are widely used. However, as a rule, a simulation model is supplemented by numerical methods. In particular, this happens when it is necessary to model relativistic and quantum effects.

We consider a semantic network to be an alternative method (regarding mathematical modelling) to represent the model of the object being studied. The use of numerical methods in simulation models destroys the integrity of the simulation model. Thus, the use of numerical methods can lead to modelling mistakes.

However, if we take the semantic network as a basis and replace some fragments of the model with calculations, we will have a more reliable method. Numerical methods can significantly reduce the dimensions of the problem. Therefore, we consider it important to create non-numerical models.

## 5. Conclusion

This study, has presented a simulation model of spacetime. The simulation (OOS, ABS) model of spacetime for SRT may be built as a special network. The network consists of time and space nodes. The main concept of the model is the synchronization of laboratory time and of the local time of cells of physical space.

Experiments that demonstrate kinematic and dynamic relativistic effects have been presented.

We represent simulation software and software documentation in the repository [6] as an open-source project. The reproducibility of experimental results can be verified. The repository will be updated with new experiments.